\documentclass[iop]{emulateapj}

\makeatletter
\renewcommand{\bibstyle@aas}{\bibpunct{(}{)}{;}{a}{,}{,}}%
\makeatother
\shorttitle{Discovery and Validation of a High-Density sub-Neptune from the K2 Mission}
\shortauthors{Espinoza et al. 2016}
\usepackage{threeparttable}

\begin{document}


\title{Discovery and Validation of a High-Density sub-Neptune from the K2 Mission}


\author{N\'estor Espinoza\altaffilmark{1,2}, 
Rafael Brahm\altaffilmark{1,2}, 
Andr\'es Jord\'an\altaffilmark{1,2}, 
James S. Jenkins\altaffilmark{3}, 
Felipe Rojas\altaffilmark{1}, 
Paula Jofr\'e\altaffilmark{4}, 
Thomas M\"adler\altaffilmark{4}, 
Markus Rabus\altaffilmark{1,5}, 
Julio Chanam\'e\altaffilmark{1,2},
Blake Pantoja\altaffilmark{3}, 
Maritza G. Soto\altaffilmark{3},
Katie M. Morzinski\altaffilmark{6},
Jared R. Males\altaffilmark{6},
Kimberly Ward-Duong\altaffilmark{7},
Laird M. Close\altaffilmark{6}
}


\altaffiltext{1}{Instituto de Astrof\'isica, Facultad de F\'isica,
    Pontificia Universidad Cat\'olica de Chile, Av.\ Vicu\~na Mackenna
    4860, 782-0436 Macul, Santiago, Chile}
    
\altaffiltext{2}{Millennium Institute of Astrophysics, Av.\ Vicu\~na Mackenna
    4860, 782-0436 Macul, Santiago, Chile}
    
\altaffiltext{3}{Departamento de Astronom\'ia, Universidad de Chile, 
Camino al Observatorio, Cerro Cal\'an, Santiago, Chile}

\altaffiltext{4}{Institute of Astronomy, University of Cambridge}

\altaffiltext{5}{Max Planck Institute for Astronomy, Heidelberg, Germany}

\altaffiltext{6}{Steward Observatory, University of Arizona, 933 N. Cherry Ave, Tucson, 
AZ 85721-0065 USA}

\altaffiltext{7}{School of Earth and Space Exploration, Arizona State University, Tempe, AZ, 85287, USA}

\begin{abstract}
We report the discovery of BD+20594b, a high density sub-Neptune exoplanet, made using 
photometry from Campaign 4 of the two-wheeled Kepler (K2) mission, ground-based radial 
velocity follow-up from HARPS and high resolution lucky and adaptive optics imaging obtained using 
AstraLux and MagAO, respectively. The host star is a bright ($V=11.04$, $K_s = 9.37$), slightly 
metal poor ([Fe/H]$=-0.15\pm 0.05$ dex) solar analogue located at $152.1^{+9.7}_{-7.4}$ pc from 
Earth, for which we find a radius of $R_*=0.928^{+0.055}_{-0.040}R_\Sun$ and a mass of 
$M_* = 0.961^{+0.032}_{-0.029}M_\Sun$. A joint analysis of the K2 photometry and HARPS radial 
velocities reveal that the planet is in a $\approx 42$ day orbit around its host star, has a radius of 
$2.23^{+0.14}_{-0.11}R_\Earth$, and a mass of $16.3^{+6.0}_{-6.1}M_\Earth$. Although the data at 
hand puts the planet in the region of the mass-radius diagram where we could expect planets with 
a pure rock (i.e. magnesium silicate) composition using two-layer models (i.e., between rock/iron 
and rock/ice compositions), we discuss more realistic three-layer composition models which can explain 
the high density of the discovered exoplanet. The fact that the planet lies in the boundary between 
``possibly rocky" and ``non-rocky" exoplanets, makes it an interesting planet for future RV follow-up.
\end{abstract}


\keywords{kepler, exoplanets}



\section{Introduction}

Since the discovery of the first rocky exoplanet \cite[term we use here to refer to planets with 
masses and radii consistent with MgSiO$_3$ and Fe compositions following][]{Rogers2015}, 
CoRoT-7b \citep{leger09,queloz09}, effort has been made to find and study the formation, composition and evolution of these 
systems, since they resemble Earth in many ways. As most rocky planets are smaller 
in size than $1.6R_\Earth$, which correspond to masses of $6M_\Earth$ \citep{WM14, wolfgang15, Rogers2015}, the 
discovery of those type of exoplanets is difficult due to the small signals that these radii and masses imply. In fact, in addition to CoRoT-7b, 
only 9 planets with secure masses and radii (i.e., masses and radii with values more than $3-\sigma$ away from zero) in 
this rocky regime exist to date: GJ1132 \citep{bt2015}, Kepler-36b \citep{carter2012}, K2-3d \citep{crossfield15,almenara15}, 
Kepler-93b \citep{dressing2015}, Kepler-10b \citep{dumusque2015,weiss2016}, Kepler-23b \citep{ford2012,HL2014}, 
Kepler-20b \citep{fressin2012}, Kepler-406b \citep{marcy2014}, and Kepler-78b \citep{so13,howard13,pepe2013,grunblatt15}. All of these 
planets have radii smaller than $\sim 1.6R_\Earth$, as has been empirically determined.

Although the sample of rocky planets is small, some interesting relationships suggest that some of these 
rocky planets might have common properties \citep{WM14}. Perhaps one of the most interesting relations was 
recently introduced by \cite{dressing2015} which, considering the planets with radii and mass 
measurements measured to better than $20\%$ precision, show that the planets follow a common 
iso-composition curve on the mass-radius diagram, along with Earth and Venus. This relation was recently 
revised by \cite{zs2016} to be a 74\% rock and 26\% Fe composition. This suggests that these small, rocky 
analogs of Earth might have similar compositions with small intrinsic scatter.

Here we report what could be a possible interesting addition to the picture of rocky worlds described above: 
a $2.23R_\Earth$ exoplanet that falls just where a pure rock (i.e., magnesium silicate) 
composition is expected in the mass radius diagram using two-layer models. Although this does not 
mean the planet has exactly this composition, its position on the diagram does makes it interesting due the fact 
that this has been used in previous works to divide the ``non-rocky" and ``possibly rocky" planets \citep{Rogers2015}. 
The discovery is made in the context of a Chilean based effort whose aim is to follow-up planetary candidates selected using data from the two-wheeled 
\textit{Kepler} (K2) mission. K2 has proven to be very effective in the search for exoplanets, enabling a plethora of new 
discoveries of planets of different sizes, which are especially interesting due to the presence of several bright host stars in 
the sample that allow detailed follow-up characterisation \citep[see, e.g., ][]{armstrong15, becker15, crossfield15, petigura2015,so15, vanderburg15}. 

The paper is structured as follows. In \S2 we present the data, which includes the K2 photometry, archival, 
new, adaptive optics (AO) and lucky imaging of the target star, along with high resolution spectra and radial velocities obtained 
with the HARPS spectrograph. \S3 presents a joint analysis of the data and presents the derived parameters of the 
planetary system. We discuss the results in \S4 and present our conclusions in \S5.

\section{Data}
\subsection{K2 Photometry}

\begin{figure*}
\plotone{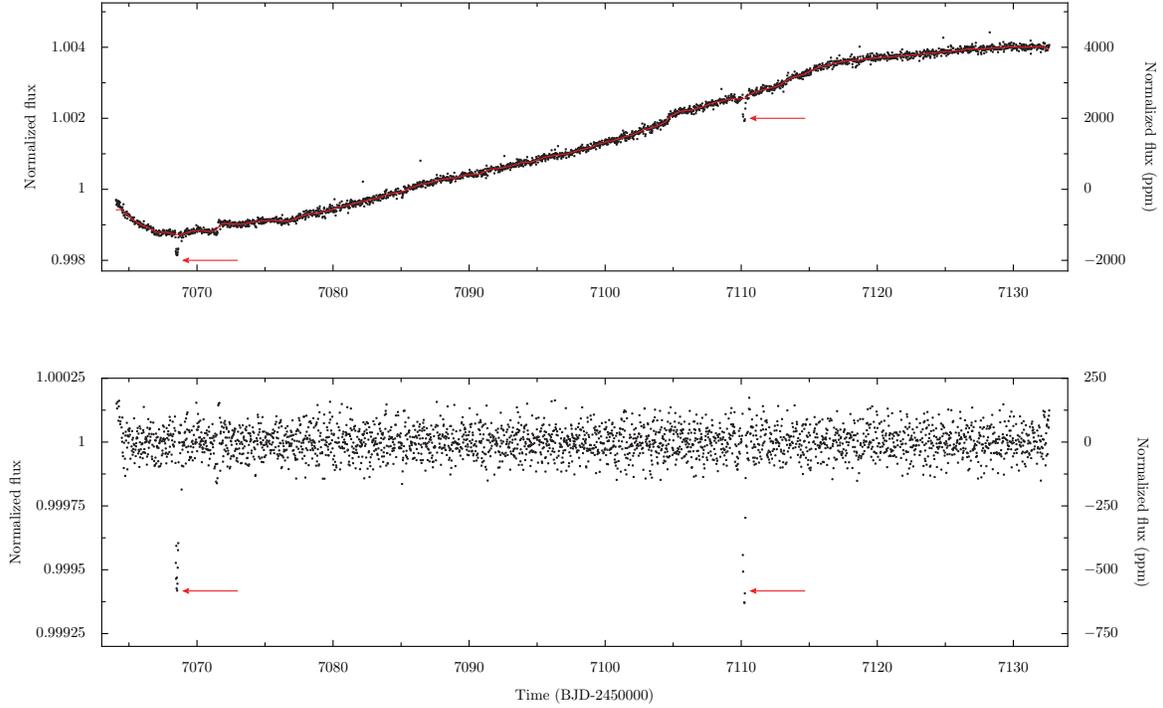}
\caption{K2 photometry \cite[obtained from][upper panel]{VJ14} and long-term and outlier corrected version of the photometry (lower panel). 
The smooth, long-term variation observed in the original photometry was removed by a smoothed median-filter, depicted in the upper panel by 
a red solid line, which was used for outlier removal (see text). Two clear transit-like events can be seen on both versions of the photometry close 
to $2457070$ and $2457110$ BJD (indicated with red arrows). Note that the  precision obtained for this lightcurve is 
$\sim 55$ ppm (rms) per point.
 \label{k2lc}}
\end{figure*}

K2 photometry for our target was obtained by the \textit{Kepler} spacecraft during Campaign 4. This field was observed
between February and April 2015 and  the data was released on September of the 
same year. We obtained the decorrelated versions of all the lightcurves in the campaign which were made publicly available for download 
by \cite{VJ14}, using the photometry with the optimal aperture, which in the case of our target star corresponded to a $\approx 3$ pixel 
radius around the target, or an aperture of $\approx 12\arcsec$ radius. We performed a transit search using 
a Box Least Squares \citep[BLS,][]{bls2002} algorithm. Once a periodic signal is detected along with 
the best-fit depth, the transit event is flagged as a pontential planetary candidate if (1) the depth is at least 
$3\sigma$ larger than the average noise level of the lightcurve (denoted by $\sigma$) and (2) if there are three or more transit events. 
Initially, because of the last requirement, the lightcurve of the target star was not flagged by our transit search pipeline. 
However, we also performed visual inspection of all the lightcurves, revealing this interesting candidate. In order to double check 
that this was indeed an astrophysical signal and not a spurious signal arising from the decorrelation method used to obtain the lightcurve, 
we also inspected the detrended lightcurves released by the \textit{Kepler} team using the PDC-MAP algorithm \citep{stumpe2012}, and the same 
signal was observed at the exact same times as the signals observed in the \cite{VJ14} photometry. We were thus confident that the signal is of astrophysical origin and proceeded to analyse the light curve.

A median filter with a 41 point ($\sim 20.5$ hour) window was used in order to further filter long-term variations of this target. The resulting median filter was 
smoothed using a Gaussian filter with a 5-point standard-deviation, and this smoothed light curve was used to normalize the light curve. Using this normalized lightcurve, 
an initial fit using our transit-fitting pipeline (see below) revealed a $P=41.7d$ period for this candidate and a lightcurve whose shape resembled that 
of a planetary transit, with a transit duration consistent with that of a planetary companion. Using the 
parameters obtained from this initial fit, we removed outliers from the out-of-transit data, discarding any points 
deviating more than 3-$\sigma$ from the median flux. The resulting normalized version of this lightcurve 
is shown on Figure~\ref{k2lc}. No other significant signals were found in the photometry.

\subsection{Reconnaissance spectroscopy}

A high resolution spectrum of this target was taken on October 21st with the CORALIE spectrograph mounted on the 
1.2m Euler Telescope in La Silla Observatory in order to obtain rough spectral parameters of the stellar host, 
and define whether this was a giant or a dwarf star. Data were reduced and analyzed using the procedures decribed in \citet{jordan2014}.
The analysis of the CORALIE spectra gave $T_\textnormal{eff}=5600$ K, $\log(g) = 4.4$ dex, 
$[\textnormal{Fe/H}]=0.0$ dex and $v\sin(i)=2.5$ km/s, which revealed that the star was a dwarf solar-type star. In 
addition, no secondary peak was seen on the cross-correlation function indicating no detectable spectroscopic binary. Because of this, 
the target was promoted to our list of planetary 
candidates despite the lack of high resolution imaging needed to rule out potential blend events.

\subsection{High precision radial velocities with HARPS}

High-precision radial velocities (RVs) were obtained from the HARPS spectrograph mounted on the 3.6m telescope at La 
Silla between October and December of 2015 in order to measure the reflex motion of the star due to the hypothetical 
planet producing the transit signal. The observations covered our predicted negative and positive quadratures, along with 
epochs in between, in order to probe possible long-term trends in the RVs indicative of a possible massive companion. 23 
spectra were taken in total with the simultaneous Thorium-Argon mode; the HARPS pipeline (DRS, version 3.8) was used to 
reduce these spectra and to obtain the (drift-corrected) radial velocities, which are calculated via cross-correlation with a G2V 
mask which is appropiate for the stellar type of the host (see \S3.1). The typical precision was $\sim 3$ m/s for each individual 
RV measurement. For each spectra, the bisector span, $S$-index, and the integrated flux of the H$_\alpha$ and \ion{He}{1} 
lines were obtained to monitor the activity of the host star and study its influence on the RVs \citep{santos2010,jenkins2011}. 
The measured RVs, along with these various calculated activity indicators, are given in Table~\ref{table:rv_list}. Although the 
times are given in UTC, they were converted to TBD (which is the time scale used by Kepler) for our joint analysis, 
which we describe in \S3.2.

\begin{deluxetable*}{ccccccccccc}[!ht]
 \tablecaption{Radial velocities obtained with the HARPS spectrograph along with various activity indicators.}
\tablehead{
     \colhead{BJD} &  RV & $\sigma_{\rm RV}$ & 
     BIS  & $\sigma_{\rm BIS}$  & $S_{H,K}$ & $\sigma_{S_{H,K}}$ 
     & ${\textnormal{H}_\alpha}$  & $\sigma_{H_\alpha}$ 
     &  \ion{He}{1} & $\sigma_{H_\alpha}$ \\
     \colhead{(UTC)} & \colhead{m sec$^{-1}$} & \colhead{m sec$^{-1}$}
     & \colhead{m sec$^{-1}$} & \colhead{m sec$^{-1}$} 
     & \colhead{dex} & \colhead{dex}
     & \colhead{dex} & \colhead{dex}
     & \colhead{dex} & \colhead{dex}
}
\startdata
$ 2457329.63450 $&$ -20333.9 $&$ 4.4 $&$ 35.0 $&$ 6.2 $&$ 0.1748 $&$ 0.0063 $&$ 0.10151 $&$ 0.00013 $&$ 0.50230 $&$ 0.00081 $\\
$ 2457329.67362 $&$ -20340.1 $&$ 3.6 $&$ 28.9 $&$ 5.0 $&$ 0.1535 $&$ 0.0050 $&$ 0.10337 $&$ 0.00013 $&$ 0.50279 $&$ 0.00081 $\\
$ 2457329.72375 $&$ -20337.9 $&$ 3.9 $&$ 34.2 $&$ 5.6 $&$ 0.1864 $&$ 0.0053 $&$ 0.10367 $&$ 0.00013 $&$ 0.50146 $&$ 0.00081 $\\
$ 2457330.80181 $&$ -20343.1 $&$ 2.6 $&$ 22.6 $&$ 3.7 $&$ 0.1483 $&$ 0.0035 $&$ 0.10167 $&$ 0.00013 $&$ 0.50787 $&$ 0.00082 $\\
$ 2457331.63418 $&$ -20342.5 $&$ 2.4 $&$ 23.2 $&$ 3.4 $&$ 0.1551 $&$ 0.0029 $&$ 0.10171 $&$ 0.00013 $&$ 0.51233 $&$ 0.00083 $\\
$ 2457331.68695 $&$ -20338.4 $&$ 2.0 $&$ 11.0 $&$ 2.8 $&$ 0.1573 $&$ 0.0026 $&$ 0.10209 $&$ 0.00013 $&$ 0.50301 $&$ 0.00081 $\\
$ 2457332.64705 $&$ -20335.7 $&$ 2.6 $&$ 20.2 $&$ 3.7 $&$ 0.1549 $&$ 0.0038 $&$ 0.10236 $&$ 0.00013 $&$ 0.50221 $&$ 0.00081 $\\
$ 2457332.72713 $&$ -20338.4 $&$ 2.0 $&$ 12.9 $&$ 2.9 $&$ 0.1459 $&$ 0.0025 $&$ 0.10178 $&$ 0.00013 $&$ 0.50273 $&$ 0.00081 $\\
$ 2457336.65528 $&$ -20345.3 $&$ 3.2 $&$ 20.8 $&$ 4.5 $&$ 0.1701 $&$ 0.0042 $&$ 0.10397 $&$ 0.00013 $&$ 0.49984 $&$ 0.00081 $\\
$ 2457336.73328 $&$ -20339.8 $&$ 3.6 $&$ 9.5 $&$ 5.1 $&$ 0.1547 $&$ 0.0047 $&$ 0.10187 $&$ 0.00013 $&$ 0.49884 $&$ 0.00081 $\\
$ 2457339.70924 $&$ -20343.1 $&$ 4.9 $&$ 14.2 $&$ 6.9 $&$ 0.1985 $&$ 0.0068 $&$ 0.09939 $&$ 0.00013 $&$ 0.49982 $&$ 0.00081 $\\
$ 2457339.72063 $&$ -20339.2 $&$ 4.1 $&$ 31.1 $&$ 5.8 $&$ 0.1738 $&$ 0.0058 $&$ 0.10496 $&$ 0.00013 $&$ 0.50575 $&$ 0.00082 $\\
$ 2457340.69354 $&$ -20340.0 $&$ 3.7 $&$ 5.9 $&$ 5.3 $&$ 0.1833 $&$ 0.0056 $&$ 0.10217 $&$ 0.00013 $&$ 0.51486 $&$ 0.00083 $\\
$ 2457340.70475 $&$ -20336.6 $&$ 3.1 $&$ 23.9 $&$ 4.3 $&$ 0.1687 $&$ 0.0044 $&$ 0.10083 $&$ 0.00013 $&$ 0.50038 $&$ 0.00081 $\\
$ 2457341.74523 $&$ -20336.6 $&$ 3.3 $&$ 18.7 $&$ 4.7 $&$ 0.1658 $&$ 0.0043 $&$ 0.10536 $&$ 0.00013 $&$ 0.50213 $&$ 0.00081 $\\
$ 2457341.75600 $&$ -20335.3 $&$ 3.2 $&$ 25.4 $&$ 4.5 $&$ 0.1491 $&$ 0.0043 $&$ 0.10274 $&$ 0.00013 $&$ 0.50658 $&$ 0.00082 $\\
$ 2457348.80101 $&$ -20330.3 $&$ 3.1 $&$ 21.6 $&$ 4.4 $&$ 0.1858 $&$ 0.0056 $&$ 0.10339 $&$ 0.00013 $&$ 0.50277 $&$ 0.00081 $\\
$ 2457360.62435 $&$ -20337.2 $&$ 2.2 $&$ 14.8 $&$ 3.2 $&$ 0.1581 $&$ 0.0029 $&$ 0.10354 $&$ 0.00013 $&$ 0.50080 $&$ 0.00081 $\\
$ 2457360.63915 $&$ -20336.8 $&$ 2.1 $&$ 11.0 $&$ 3.0 $&$ 0.1585 $&$ 0.0027 $&$ 0.10237 $&$ 0.00013 $&$ 0.50273 $&$ 0.00081 $\\
$ 2457361.66418 $&$ -20337.3 $&$ 3.1 $&$ 30.7 $&$ 4.4 $&$ 0.1719 $&$ 0.0042 $&$ 0.10607 $&$ 0.00013 $&$ 0.50020 $&$ 0.00081 $\\
$ 2457361.67814 $&$ -20337.0 $&$ 2.9 $&$ 32.8 $&$ 4.1 $&$ 0.1533 $&$ 0.0038 $&$ 0.10231 $&$ 0.00013 $&$ 0.49660 $&$ 0.00080 $\\
$ 2457362.66191 $&$ -20331.0 $&$ 2.2 $&$ 15.4 $&$ 3.1 $&$ 0.1517 $&$ 0.0031 $&$ 0.10313 $&$ 0.00013 $&$ 0.49866 $&$ 0.00081 $\\
$ 2457362.67602 $&$ -20335.6 $&$ 2.2 $&$ 13.5 $&$ 3.1 $&$ 0.1575 $&$ 0.0032 $&$ 0.10479 $&$ 0.00013 $&$ 0.50121 $&$ 0.00081 $
\enddata
 \label{table:rv_list}
\end{deluxetable*}

\subsection{Archival and New Imaging}

\begin{figure*}
\plottwo{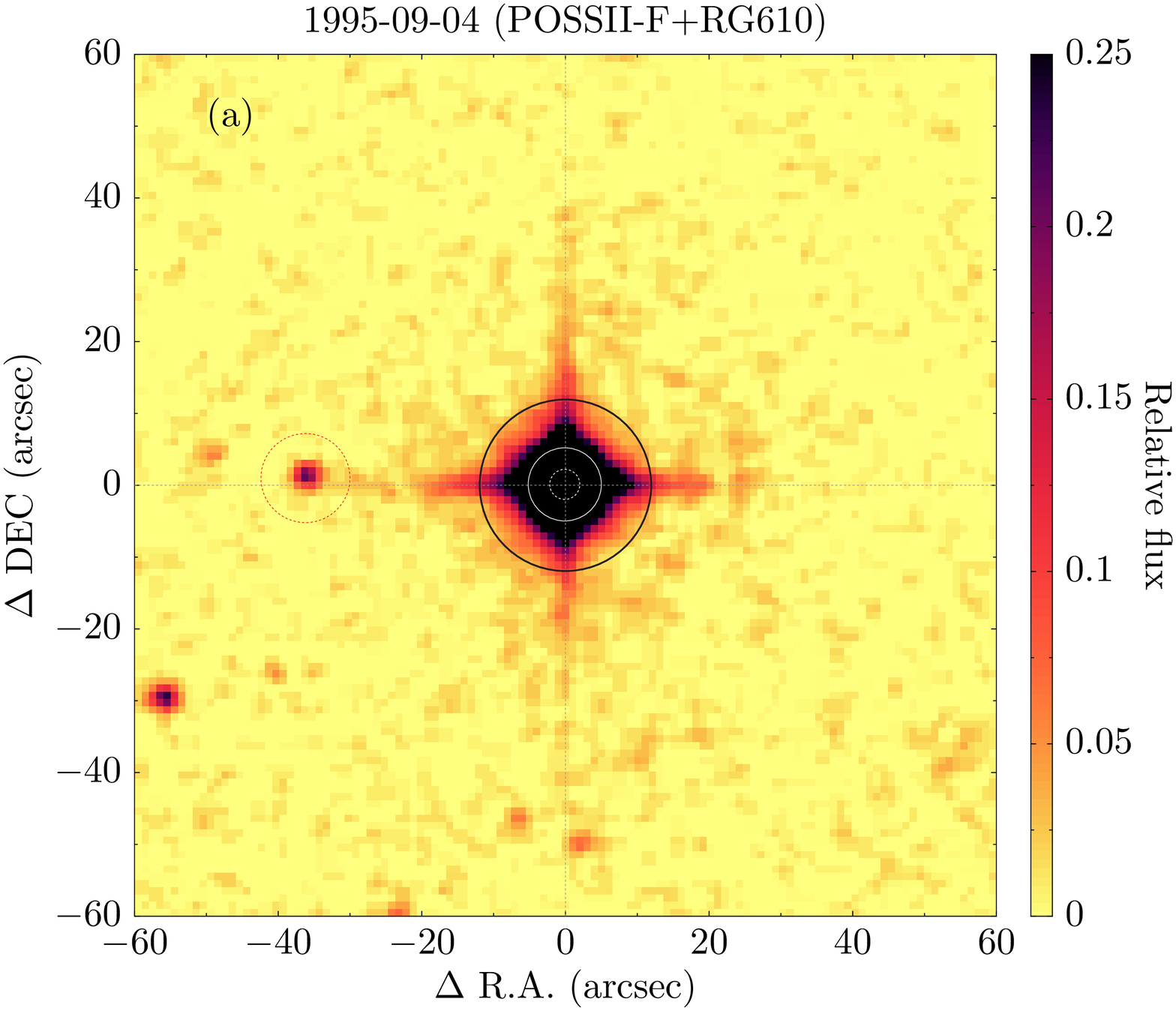}{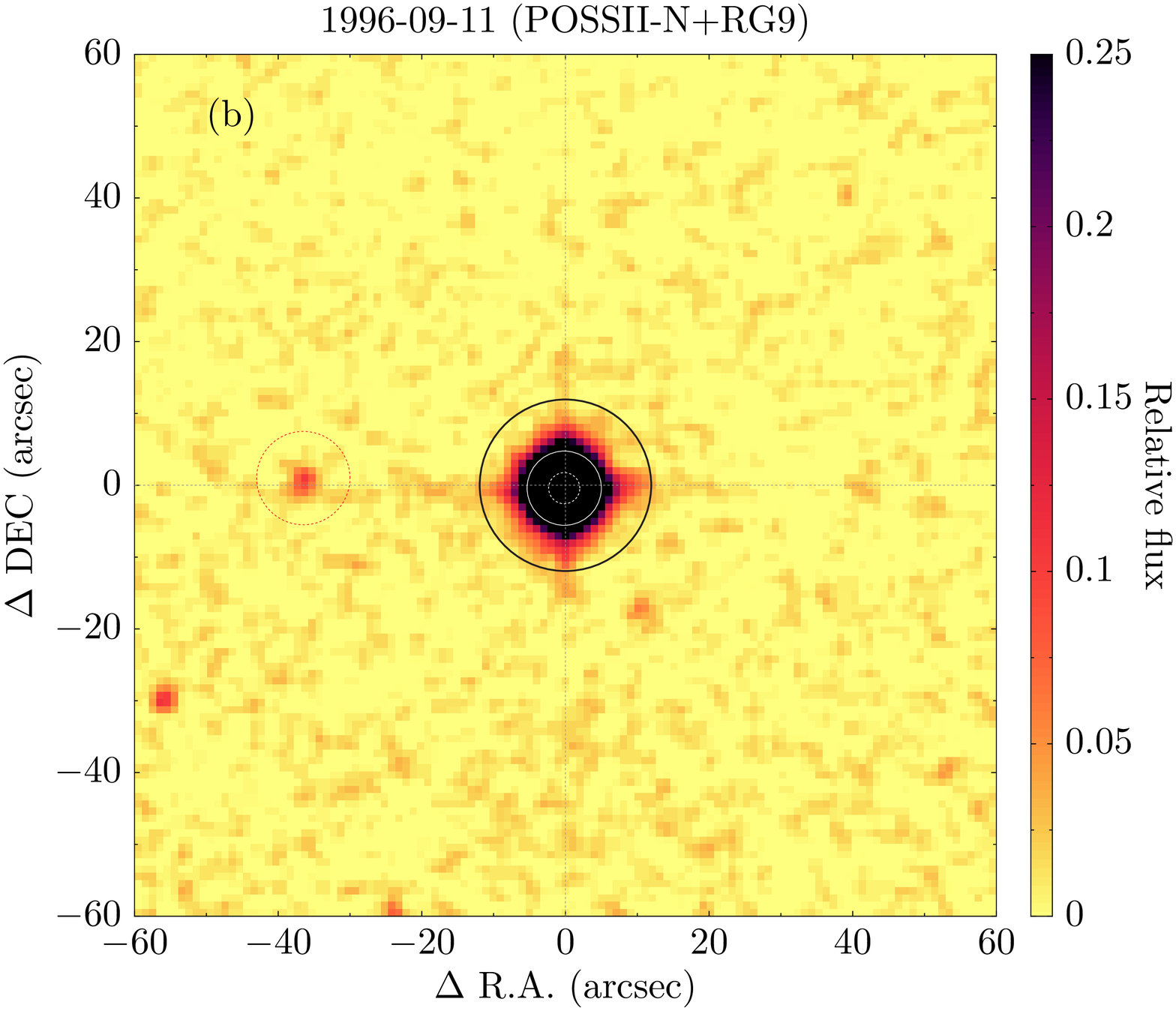}
\caption{Archival imaging for our target at the coordinates given in the EPIC catalog obtained with different versions of POSS: (a) POSSII-F survey, taken 
with a red filter and (b) POSSII-N survey, taken with an infrared filter. The black circle indicates the aperture used for our K2 data. The white solid circle has a 
radius of $5''$ and the dashed circle a radius of $2''$ for illustration purposes; these are centered on the measured centroid of the target star. The red circle 
to the left of the target star marks an object which is $\sim 8.2$ magnitudes fainter than the target in $R$ (see text).
 \label{archival-image}}
\end{figure*}

Archival imaging was obtained from the STScI Digitized Sky Survey\footnote{\url{http://stdatu.stsci.edu/cgi-bin/dss\_form}} at the EPIC coordinates 
of our target. Data are from the Palomar Observatory Sky Survey (POSS). In Figure~\ref{archival-image} we show the best images among the available 
archival images in terms of the measured FWHM. We show images taken at two epochs and with two filters: one obtained in 1995 using the RG610 
filter (red\footnote{Transmission curve available at \url{http://www.cadc-ccda.hia-iha.nrc-cnrc.gc.ca/en/dss/TransmissionCurves/POSSII-F-IIIaF-RG610.txt}}, 
$590-715$ nm), taken by the POSSII-F and one using the RG9 filter (near-infrarred\footnote{Transmission curve available at 
\url{http://www.cadc-ccda.hia-iha.nrc-cnrc.gc.ca/en/dss/TransmissionCurves/POSSII-N-IVN-RG9.txt}}, $700-970$ nm) obtained in 1996 by the POSSII-N. For 
reference, we show the aperture used to obtain our K2 photometry (black circle, $12\arcsec$) along with circles with $5\arcsec$ (white solid line) and $2\arcsec$ (white dashed 
line) radii which are centered on the centroid of our target star, which was obtained by fitting a 2D Gaussian to the intensity profile. 

\begin{figure}
\plotone{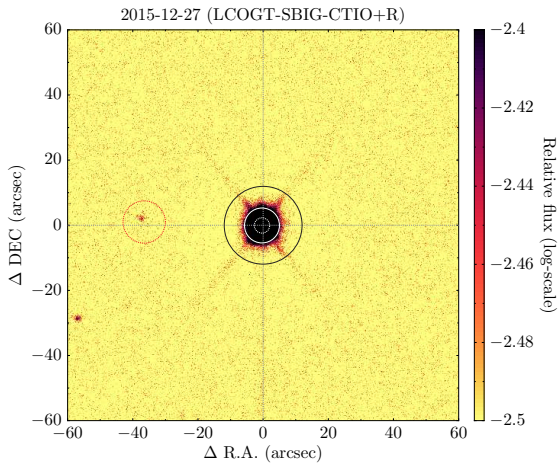}
\caption{Modern imaging of our target obtained from LOCGT from CTIO using the SBIG camera with the $R$ filter on 2015/12/27. Note that, although the 
circles have the same meanings as the ones in Figure~\ref{archival-image}, the scale here is different.
 \label{modern-image}}
\end{figure}

New imaging was obtained using the Las Cumbres Observatory Global Telescope Network (LCOGT). Four images were taken using the SBIG camera with the 
Bessel $R$ filter on UT 2015/12/27 from the Cerro Tololo Interamerican Observatory (CTIO). Our target star reached close-to saturation counts 
($\sim47000$ counts) in order to have enough photons to observe the close-by stars present in the POSS images. Figure~\ref{modern-image} shows the 
resulting image obtained by median-combining our four images, along with the same circles as those drawn on Figure~\ref{archival-image}. 

Given that the largest potential source of false-positive detections in our case comes from blended eclipsing binary systems mimicking a planetary transit event, we note 
that, given that the depth of the observed transit is $\sim 0.05\%$, if a blended eclipsing binary system was responsible of the observed depth, then assuming a 
total eclipse of the primary (which is the worst case scenario; all other scenarios should be easier to detect), the eclipsed star would have to be $\sim 8.23$ 
magnitudes fainter than our target star in the {\em Kepler} bandpass. We can confidently rule out such a bright star down to a distance of $9\arcsec$ of the target star 
with the POSSII and LCOGT images. For reference, the closest star to the left of the target star (indicated with a red circle) in Figures \ref{archival-image} and 
\ref{modern-image} is $\sim 8.2$ magnitudes fainter than the target star in the $R$ band. As can be seen on the images, a star that bright would be evident in the 
archival POSS images and/or on our new LCOGT images at distances larger than $9\arcsec$.

\subsection{Adaptive optics \& lucky imaging}
\begin{figure*}
\plottwo{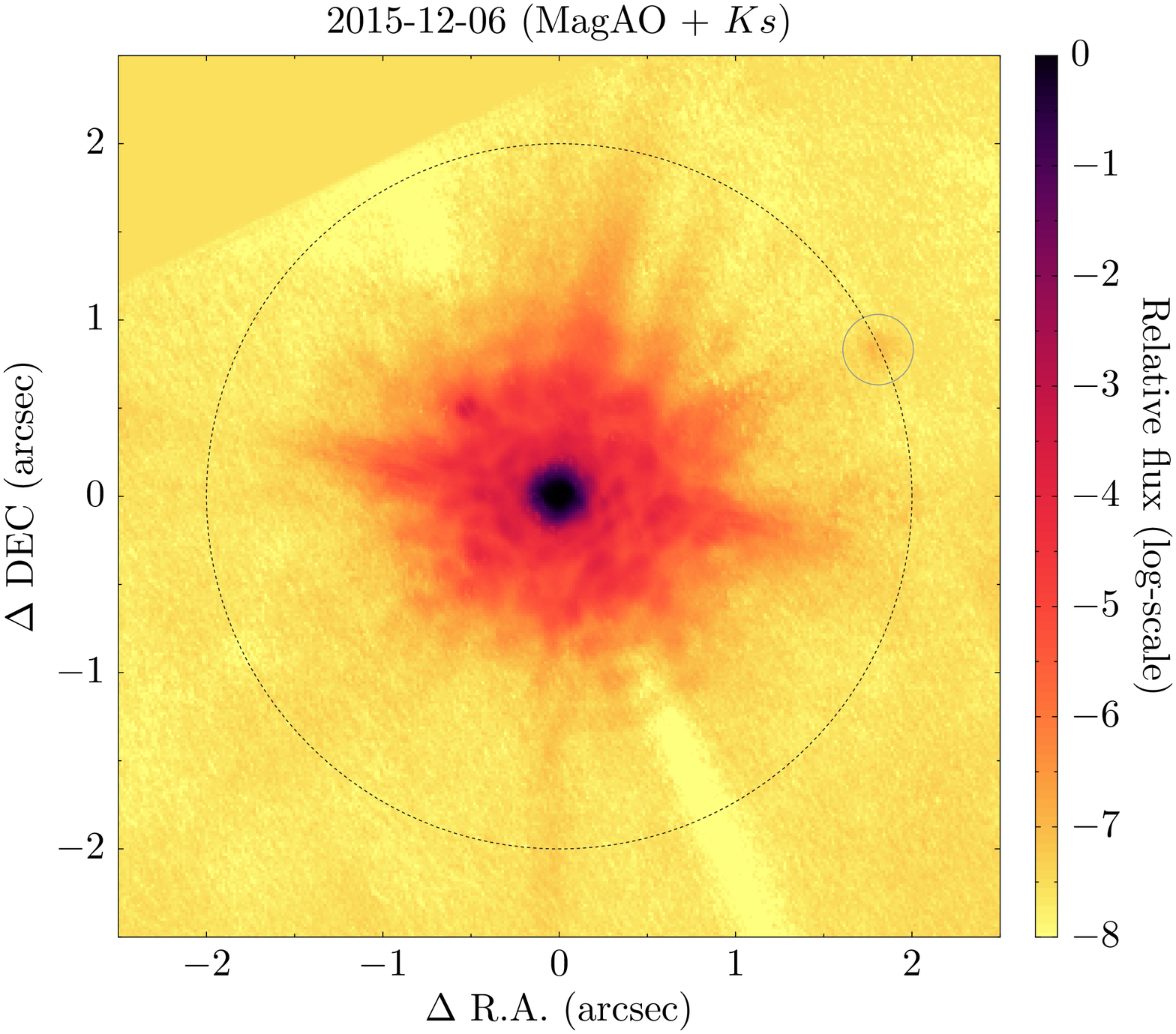}{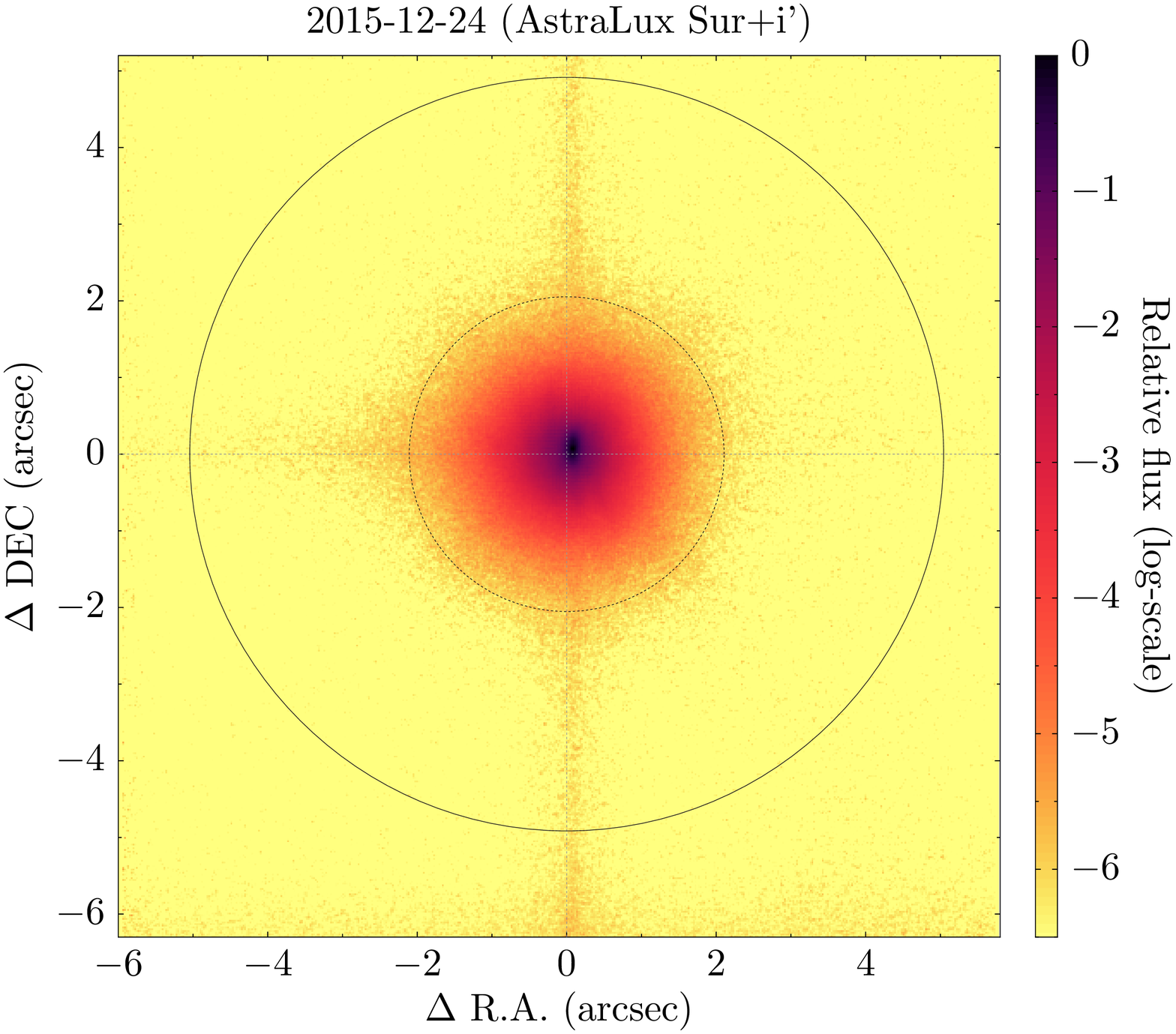}
\caption{(\textit{Left}) Adaptive optics image (log-scale) obtained with MagAO+Clio2 on 2015-12-06. The black dashed circle has 
a $2''$ radius for illustration and comparison with Figure \ref{archival-image}; the grey circle mark a faint source found on our 
image, which was above our contrast limit but we identified as being of instrumental origin (see text). (\textit{Right}) AstraLux Sur $i'$-band 
observations of our candidate on 2015-12-24. The inner black dashed circle indicates $2''$, while the outer black solid circle indicates $5''$ for comparison with Figure \ref{archival-image}. 
 \label{ao-lucky-image}}
\end{figure*}

Adaptive optics (AO) imaging was obtained using MagAO+Clio2 instrument mounted at the Magellan Clay telescope in Las Campanas Observatory on 
December 6th using the $K_s$ filter with the full Clio2 $1024\times512$ pixel frames of the narrow camera (f/37.7). The natural guide star system 
was used and, because our target is relatively bright, it was used as the guide star. 32 images with exposure times of 30 sec each were taken in five 
different positions of the camera (nodding), all of them at different rotator offset angles. Due to a motor failure of the instrument, the nodding and 
rotation patterns were not able to cover the full $16\arcsec\times8\arcsec$ field of view around the star. However, it gave us enough data to rule 
out stars within a $2\arcsec$ radius. We follow methods similar to those described in \cite{morzinski15} to reduce our images, which we briefly describe 
here; a \texttt{Python} implementation of such methods is available at Github\footnote{\url{https://github.com/nespinoza/ao-reduction}}. First, the images 
were corrected by dark current but not flat fielded, because the flats show an uneven flux level as a result of optical distortions and not of intrinsic pixel sensitivities 
\citep[see section A.3 in ][for a detailed explanation of this effect]{morzinski15}. A bad pixel mask provided by \cite{morzinski15} was used in order to mask 
bad pixels. After these corrections are applied to each image, we obtain a median image using our 32 frames in order to get an estimate of the 
background flux, which we then subtract from each of the individual frames. In order to further correct for differences in the sky backgrounds of each image, 
we apply a 2D median filter with a 200 pixel ($\approx 3\arcsec$) window which takes care of large-scale fluctuations of each image. The background-subtracted 
images are then merged by first rotating them to the true north \citep[using the astrometric calibration described in][]{morzinski15} and combined using the centroid 
of our target star (obtained by fitting a 2D Gaussian to the profile) as a common reference point between the images. Our resulting AO image, obtained by combining 
our 32 images, is shown in Figure~\ref{ao-lucky-image}. A 2D gaussian fit to the target star gives a FWHM of $0\farcs2$, which we set as our resolution limit.

\begin{figure}
\plotone{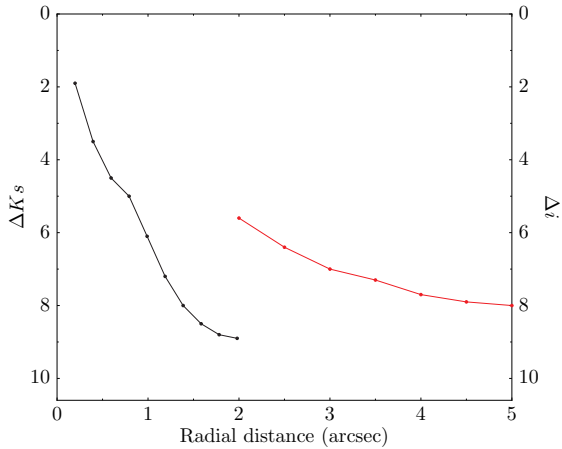}
\caption{5-$\sigma$ contrast curves obtained from our MagAO+Clio2 $K_s$ band (black line) and AstraLux Sur $i'$-band (red line) observations of our candidate.
 \label{contrast-plot}}
\end{figure}

The limiting contrasts in our AO observations in the $K_s$ band were estimated as follows. First, a 2D gaussian fit to the target star was made and used to 
remove it from the image. Although a 2D gaussian does not perform a perfect fit at the center, the fit is good enough for the wings of the PSF, which is our aim. 
Then, at each radial distance $n\times$FWHM away from the target star, where $n=1,2,...,15$ is an integer, a fake source was injected 
at $15$ different angles. Sources with magnitude differences from 11 to 0 were injected in $0.1$ steps, and a detection was defined if 3 or more pixels were 
5-sigma above the median flux level at that position. The results of our injection and recovery experiments are plotted in Figure~\ref{contrast-plot}.

Only one source was detected at $\sim 2\arcsec$ from the target. The shape and position of this object is inconsistent with a speckle but is very faint: we measure a 
magnitude difference of $\Delta K_s = 9.8$ with the target and is thus just above our contrast level at that position (see Figure~\ref{ao-lucky-image}, the source is 
indicated with a grey circle in the upper right). A careful assessment of the PSF shape, however, made it inconsistent with the object having the same PSF shape 
as our star. Comparing its PSF with known ``ghosts" on the image, on the other hand, revealed that this source is not of astrophysical but of instrumental origin.

In order to search for companions at larger separations, lucky imaging was obtained with AstraLux Sur mounted on the New Technology Telescope  
(NTT) at La Silla Observatory \citep{hippler09} on 2015/12/24 using the $i'$ band. Figure~\ref{ao-lucky-image} shows our final image obtained by combining the best 
10\% images with a drizzle algorithm. Because the PSF shape obtained for our lucky imaging is complex and we already ruled out companions inside a $2\arcsec$ radius 
with Magellan+Clio2, and given that our objective with lucky imaging was to rule out companions at larger angular distances, we did not perform PSF substraction algorithms 
in order to obtain the $5-\sigma$ contrasts at those distances. Instead, we used simple aperture photometry in order to estimate the $5-\sigma$ contrasts outside the 
$2\arcsec$ radius by performing a procedure similar to that described in \cite{wollert15}. In summary, we estimated the noise level in a $5\times 5$ box at each radial 
distance at $15$ different angles for distances larger than $2\arcsec$ from the estimated centroid of the image (where the contribution of the target star' PSF to the 
background level is low), and calculated the magnitude contrast by obtaining the flux of the target star using a 5-pixel radius around it and a 5-pixel radius about 
the desired distance from the star, where $5-\sigma$ counts are summed to each pixel at that distance before performing the aperture photometry. Then, the 
magnitude contrast at a given distance is obtained as the average value obtained at the different angles. The resulting $5-\sigma$ contrasts are presented 
in Figure~\ref{contrast-plot}. We study the constrains that our archival, new, AO and lucky imaging put on the false-positive probabilities and transit dilutions on 
the next section.

\section{Analysis}
\subsection{Stellar properties}

\begin{table*}[!ht]
 \begin{center}
 \caption{Stellar parameters of BD+20594.}
 \label{table:stellar-params}
 \begin{threeparttable}
  \centering
  \begin{tabular}{ lcr }
   \hline
   \hline
     Parameter &  Value & Source \\
   \hline
Identifying Information\\
~~~EPIC ID & 210848071 & EPIC\\
~~~2MASS ID & 03343623+2035574 & 2MASS\\
~~~R.A. (J2000, h:m:s) & 03$^h$34$^m$36.23$s$ & EPIC\\
~~~DEC (J2000, d:m:s) & 20$^o$35$'$57.23$''$ & EPIC\\
~~~R.A. p.m. (mas/yr)  & $36.7\pm0.7$ & UCAC4\\
~~~DEC p.m. (mas/yr) & $-51.8\pm1.3$ & UCAC4\\
Spectroscopic properties\\
~~~$T_\textnormal{eff}$ (K) & $5766\pm 99$ & ZASPE\\
~~~Spectral Type & G & ZASPE\\
~~~[Fe/H] (dex) & $-0.15\pm 0.05$ & ZASPE\\
~~~$\log g_*$ (cgs)& $4.5\pm 0.08$ & ZASPE\\
~~~$v\sin(i)$ (km/s)& $3.3\pm 0.31$ & ZASPE\\
Photometric properties\\\
~~~$K_p$ (mag)& 11.04 & EPIC\\
~~~$B$ (mag)& $11.728\pm 0.044$ & APASS\\
~~~$V$ (mag)& $11.038\pm 0.047$ & APASS\\
~~~$g'$ (mag)& $11.352\pm 0.039$ & APASS\\
~~~$r'$ (mag)& $11.872\pm 0.050$ & APASS\\
~~~$i'$ (mag)& $10.918\pm 0.540$ & APASS\\
~~~$J$ (mag)& $9.770\pm 0.022$ & 2MASS\\
~~~$H$ (mag)& $9.432\pm 0.022$ & 2MASS\\
~~~$Ks$ (mag)& $9.368\pm 0.018$ & 2MASS\\
Derived properties\\
\vspace{0.1cm}
~~~$M_*$ ($M_\Sun$)& $0.961^{+0.032}_{-0.029}$ & \texttt{isochrones}+ZASPE\\
\vspace{0.1cm}
~~~$R_*$ ($R_\Sun$)& $0.928^{+0.055}_{-0.040}$ & \texttt{isochrones}+ZASPE\\
\vspace{0.1cm}
~~~$\rho_*$ (g/cm$^3$)& $1.70^{+0.20}_{-0.26}$ & \texttt{isochrones}+ZASPE\\
\vspace{0.1cm}
~~~$L_*$ ($L_\Sun$)& $0.88^{+0.15}_{-0.12}$ & \texttt{isochrones}+ZASPE\\
\vspace{0.1cm}
~~~Distance (pc)& $152.1^{+9.7}_{-7.4}$ & \texttt{isochrones}+ZASPE\\
\vspace{0.1cm}
~~~Age (Gyr)& $3.34^{+1.95}_{-1.49}$ & \texttt{isochrones}+ZASPE\\
   \hline
   \end{tabular}
      \textit{Note}. Logarithms given in base 10. 
  \end{threeparttable}
 \end{center}
 \end{table*}

In order to obtain the properties of the host star, we made use of both photometric and spectroscopic 
observables of our target. For the former, we retrieved $B$,$V$,$g$,$r$ and $i$ photometric magnitudes 
from the AAVSO Photometric All-Sky Survey \citep[APASS,][]{apass} and $J$, $H$ and $K$ 
photometric magnitudes from 2MASS for our analysis. For the spectroscopic observables, we used 
the Zonal Atmospherical Stellar Parameter Estimator \citep[\texttt{ZASPE},][]{brahm2016} algorithm 
using our HARPS spectra as input.  \texttt{ZASPE} estimates the atmospheric stellar parameters and $v \sin i$ from 
our high resolution echelle spectra via a least squares method against a grid of synthetic spectra in the most 
sensitive zones of the spectra to changes in the atmospheric parameters. \texttt{ZASPE} obtains reliable 
errors in the parameters, as well as the correlations between them by assuming that the principal 
source of error is the systematic mismatch between the data and the optimal synthetic spectra, which 
arises from the imperfect modelling of the stellar atmosphere or from poorly determined parameters of 
the atomic transitions. We used a synthetic grid provided by \cite{brahm2016} and the spectral region considered for the 
analysis was from 5000 $\AA$ to 6000 $\AA$, which includes a large number of atomic transitions and the 
pressure sensitive Mg Ib lines. The resulting atmospheric parameters obtained through this procedure were 
$T_{\textnormal{eff}} = 5766\pm 99$ K, $\log(g) = 4.5\pm 0.08$, $[\textnormal{Fe/H}] = -0.15\pm 0.05$ and 
$v\sin(i) = 3.3\pm 0.31$ km/s. With these spectroscopic parameters at hand and the photometric properties, 
we made use of the Dartmouth Stellar Evolution Database \citep{dotter2008} to obtain the radius, mass, age and 
distance to the host star using isochrone fitting with the \texttt{isochrones} package \citep{morton2015}. We take 
into account the uncertainties in the photometric and spectroscopic observables to estimate the stellar properties, 
using the \texttt{emcee} \citep{emcee2013} implementation of the affine invariant Markov 
Chain Monte Carlo (MCMC) ensemble sampler proposed in \cite{GW2010} in order to explore the posterior parameter space. 
We obtain a radius of $R_* = 0.928^{+0.055}_{-0.040}R_\Sun$, mass $M_* = 0.961^{+0.032}_{-0.029}M_\Sun$, 
age of $3.3^{+1.9}_{-1.5}$ Gyr and a distance to the host star of $152.1^{+9.7}_{-7.4}$ pc. The distance to the star 
was also estimated using the spectroscopic twin method described in \cite{jofre2015}, which is independent of any 
stellar models. The values obtained were $158.3 \pm 5.4$ pc when using 2MASS J band photometry and $160.0 \pm 5.7$ 
pc if H band photometry was used instead, where the stars HIP 1954, HIP 36512, HIP 49728 and HIP 58950 were 
used as reference for the parallax. Those values are in very good agreement with the value obtained from isochrone 
fitting. The stellar parameters of the host star are sumarized in Table~\ref{table:stellar-params}.

\subsection{Joint analysis}

We performed a joint analysis of the photometry and the radial velocities using the \textbf{EXO}planet tra\textbf{N}sits and 
r\textbf{A}d\textbf{I}al ve\textbf{L}ocity fitt\textbf{ER}, \texttt{exonailer}, which is made publicly available at 
Github\footnote{\url{http://www.github.com/nespinoza/exonailer}}. For the transit modeling, \texttt{exonailer} makes use of 
the \texttt{batman} code \citep{batman2015}, which allows 
the user to use different limb-darkening laws in an easy and efficient way. If chosen to be free parameters, the sampling of the limb-darkening 
coefficients is performed in an informative way using the triangular sampling technique described in \cite{kipping2013}. For the quadratic 
and square-root laws, we use the transformations described in \cite{kipping2013} in order to sample the physically plausible values of the 
limb-darkening coefficients. For the logarithmic law we use the transformations described in 
\cite{EJ2016}, which presents the sampling of the limb darkening parameters for the more usual form of the logarithmic law to allow for easier 
comparison with theoretical tables \cite[if the geometry of the system is properly taken into account, see][]{EJa2015}. The code also allows the 
user to fit the lightcurve assuming either a pure white-noise model or an underlying flicker ($1/f$) noise plus white-noise model using the 
wavelet-based technique described in \cite{CW2009}. For the RV modelling, \texttt{exonailer} assumes Gaussian uncertainties and adds a 
jitter term in quadrature to them. The joint analysis is then performed using the \texttt{emcee} MCMC ensemble sampler \citep{emcee2013}.

For the joint modelling of the dataset presented here, we tried both eccentric and circular fits. For the radial velocities, 
uninformative priors were set on the semi-amplitude, $K$, and the RV zero point, $\mu$. The former was centered on 
zero, while the latter was centered on the observed mean of the RV dataset. Note that our priors allow us to explore 
negative radial velocity amplitudes, which is intentional as we want to explore the possibility of the RVs being 
consistent with a flat line (i.e., $K=0$). Initially a jitter term was added but was fully consistent with zero, so we fixed it to zero 
in our analysis. As for the non-circular solutions, flat priors were set on $e$ and on $\omega$ instead of fitting for the 
Laplace parameters $e\cos(\omega)$ and $e\sin(\omega)$ because these imply implicit priors on the parameters that 
we want to avoid \citep{anglada-escude2013}. For the lightcurve modelling, we used the selective resampling technique 
described in \cite{kipping2010} in order to account for the 30 min cadence of the K2 photometry, which has as a 
consequence the smearing of the transit shape. In order to minimize the biases in the retrieved transit parameters 
we fit for the limb darkening coefficients in our analysis \citep[see][]{EJa2015}. In order to decide which limb-darkening 
law to use, we apply the method described in \cite{EJ2016} which, through simulations and given the lightcurves properties, 
aids in selecting the best limb-darkening law in terms of both precision and bias using a mean-squared error (MSE) approach. 
In this case, the law that provides the minimum MSE is the quadratic law, and we use this law in order to parametrize the 
limb-darkening effect. In addition, the K2 photometry is not good enough to constrain the ingress 
and egress times because only two transits were observed in long-cadence mode, which provides poor phase coverage; 
this implies that the errors on $a/R_*$ are rather large. Because of this, we took advantage of the stellar parameters obtained 
with our HARPS spectra, and derived a value for this parameter from them \cite[see ][]{sozzetti2007} of $a/R_* = 54.83^{+2.19}_{-3.16}$. 
This value was used as a prior in our joint analysis in the form of a Gaussian prior. We used the largest of the errorbars as the 
standard deviation of the distribution, which is centered on the quoted median value of the parameter\footnote{Performing a joint 
analysis with a large uniform prior on $a_/R_*$ spanning from $a/R_* \in (25,70)$ gives a posterior estimate of 
$a/R_* = 55.92^{+5.64}_{-13.11}$ for this parameter, which is in excellent agreement with this spectroscopically derived value.}. 
We tried both fitting a flicker-noise model and a white-noise model, but the flicker noise model parameters were consistent with 
no $1/f$ noise component, so the fit was finally obtained assuming white noise. $500$ \textit{walkers} were used to evolve the MCMC, 
and each one explored the parameter space in $2000$ links, $1500$ of which were used as burn-in samples. This gave a total of 
$500$ links sampled from the posterior per \textit{walker}, giving a total of $250000$ samples from the posterior distribution. These 
samples were tested to converge both visually and using the \cite{geweke92} convergence test. 

\begin{figure*}
\plotone{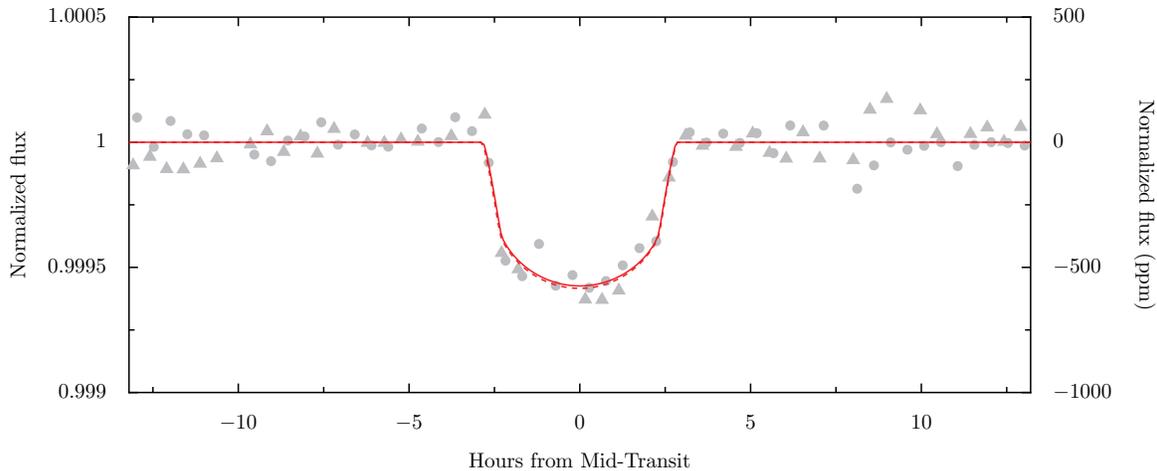}
\caption{Phase-folded photometry (grey points; circles the first transit, triangles the second transit) and best-fit transit lightcurve for the circular (red, solid line) and 
eccentric (red, dashed line) fits for our planet obtained from our joint analysis. Note that the difference in the lightcurve for both fits is very small.
 \label{k2lc-fit}}
\end{figure*}

\begin{figure}
\plotone{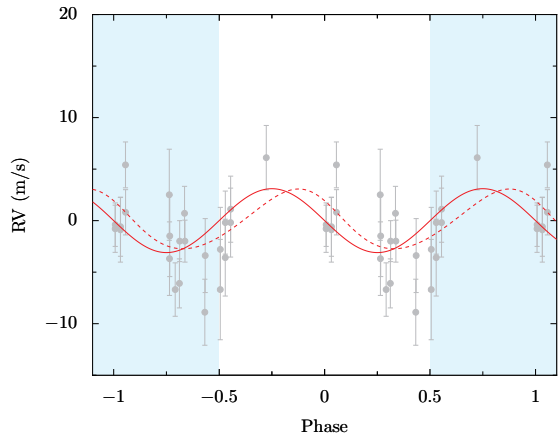}
\caption{Phase-folded HARPS radial velocities (grey) and best-fit radial velocity models for both circular (red, solid line) and eccentric 
(red, dashed line) fits using our joint analysis. The light blue bands indicate regions that have been repeated for better visualisation of the RV curve.
 \label{k2rv-fit}}
\end{figure}

Figures~\ref{k2lc-fit} and \ref{k2rv-fit} show close-ups to the phased photometry and radial velocities, respectively, along with the 
best-fit models for both circular (red, solid line) and non-circular (red, dashed line) fits obtained from our joint analysis of the dataset. 
The lightcurve fits for both models are very similar, but in the RVs the differences are evident. In particular, the eccentric fit gives rise 
to a slightly smaller semi-amplitude than (yet, consistent with) the one obtained with the circular fit. For the eccentric fit, we obtain 
$e = 0.096^{+0.089}_{-0.066}$, $\omega = 53^{+17}_{-23}$ degs and a semi-amplitude of $K = 2.9^{+1.1}_{-1.0}$ m sec$^{-1}$. For the 
circular orbit, we find a semi-amplitude of $K = 3.1^{+1.1}_{-1.1}$ m sec$^{-1}$. Since the differences on the lightcurves are very small, 
we analyze the likelihood function of the radial-velocity data in order to compare the models and decide which is preferred by the 
data. We obtain that both models are indistinguishable, with both the AIC ($\Delta\textnormal{AIC}=2$) and BIC 
($\Delta\textnormal{BIC}=2$) values being $\sim 2$. We thus choose the simpler model of those two, which is the circular model, and 
report the final parameters using this as our final model.

The resulting parameters of our fit are tabulated in Table~\ref{table:planet-params}. It is interesting to note that 
the radial velocity semi-amplitude is inconsistent with zero by almost $3\sigma$. Moreover, we are confident 
that those variations do not arise from activity as all the correlation coefficients we calculate between our RVs 
and the different activity indexes given in Table \ref{table:rv_list} give correlation coefficients which are consistent 
with $0$ at $\approx 1\sigma$, and all variations of the activity indices at the period and time of transit-center 
found for our target are consistent with flat lines. Interestingly, the radial-velocity semi-amplitude is large for 
a planetary radius of only $R_p = 2.23^{+0.14}_{-0.11}R_\Earth$; the $K=3.1^{+1.1}_{-1.1}$ m/s 
semi-amplitude implies a mass of $M_p = 16.3^{+6.0}_{-6.1} M_\Earth$, which at face value could be 
consistent with a rocky composition, a rare property for a Neptune-sized exoplanet such as BD+20594b. We 
caution, however, that this interpretation has to be taken with care, as we have poor phase coverage on the 
``up" quadrature. We put these values in the context of discovered exoplanets of similar size in \S4.

\begin{table}[!ht]
 \caption{Orbital and planetary parameters for BD+20594.}
 \label{table:planet-params}
 \begin{threeparttable}
  \centering
  \begin{tabular}{ lcl }
   \hline
   \hline
     Parameter &  Prior & Posterior Value \\
   \hline
Lightcurve parameters\\
\vspace{0.1cm}
~~~$P$ (days)\dotfill    & $\mathcal{N}(41.68,0.1)$ & 41.6855$^{+0.0030}_{-0.0031}$ \\
\vspace{0.1cm}
~~~$T_0-2450000$ (${\textnormal{BJD}_{\textnormal{TDB}}}$)\dotfill    & $\mathcal{N}(7151.90,0.1)$ & 7151.9021$^{+0.0042}_{-0.0047}$ \\
\vspace{0.1cm}
~~~$a/R_{\star}$ \dotfill    &$\mathcal{N}(54.83,3.16)$& $55.8^{+3.3}_{-3.3}$ \\
\vspace{0.1cm}
~~~$R_{p}/R_{\star}$\dotfill    & $\mathcal{U}(0,0.1)$ & 0.02204$^{+0.00058}_{-0.00057}$ \\
\vspace{0.1cm}
~~~$i$ (deg)\dotfill & $\mathcal{U}(80,90)$ &89.55$^{+0.17}_{-0.14}$\\
\vspace{0.1cm}
~~~$q_1$ \dotfill & $\mathcal{U}(0,1)$&$0.38^{+0.29}_{-0.16}$\\
\vspace{0.1cm}
~~~$q_2$ \dotfill & $\mathcal{U}(0,1)$& $0.52^{+0.32}_{-0.30}$\\
\vspace{0.1cm}
~~~$\sigma_w$ (ppm) \dotfill & $\mathcal{J}(50,80)$ & 55.00$^{+0.73}_{-0.72}$\\
\vspace{0.1cm}
RV parameters\\
\vspace{0.1cm}
~~~$K$ (m s$^{-1}$)\dotfill   & $\mathcal{N}(0,100)$ & $3.1^{+1.1}_{-1.1}$\\
\vspace{0.1cm}
~~~$\mu$ (km s$^{-1}$)\dotfill    & $\mathcal{N}(-20.337,0.1)$ & $-20.33638^{+0.00073}_{-0.00073}$ \\
\vspace{0.1cm}
~~~$e$ \dotfill    & --- & $0$ (fixed) \\
\vspace{0.1cm}
Derived Parameters\\
\vspace{0.1cm}
~~~$M_p$ ($M_\Earth$)       \dotfill      &---& $16.3^{+6.0}_{-6.1}$  \\
\vspace{0.1cm}
~~~$R_p$ ($R_\Earth$)       \dotfill       &---& $2.23^{+0.14}_{-0.11}$  \\
\vspace{0.1cm}
~~~$\rho_p$ (g/cm$^3$)       \dotfill      &---& $7.89^{+3.4}_{-3.1}$  \\
\vspace{0.1cm}
~~~$\log g_p$ (cgs)             \dotfill      &---& $3.50^{+0.14}_{-0.21}$  \\
\vspace{0.1cm}
~~~$a$ (AU)             \dotfill      &---& $0.241^{+0.019}_{-0.017}$  \\
\vspace{0.1cm}
~~~$V_\textnormal{esc}$ (km/s)             \dotfill      &---& $30.2^{+5.3}_{-6.2}$  \\
\vspace{0.1cm}
~~~$T_\textnormal{eq}$ (K)  \dotfill          &&\\
\vspace{0.1cm}
~~~\ Bond albedo of $0.0$     &---& $546^{+19}_{-18}$  \\
\vspace{0.1cm}
~~~\ Bond albedo of $0.75$        &---&     $386^{+13}_{-12}$  \\

   \hline
   \end{tabular}
   \textit{Note}. Logarithms given in base 10. $\mathcal{N}(\mu,\sigma)$ stands for a normal prior with mean $\mu$ and standard-deviation 
   $\sigma$, $\mathcal{U}(a,b)$ stands for a uniform prior with limits $a$ and $b$ and $\mathcal{J}(a,b)$ stands for a Jeffrey's prior 
   with the same limits.
  \end{threeparttable}
 \end{table}
 
\subsection{Planet scenario validation}

In order to validate the planet scenario which we have implied in the past sub-section, we make use of the formalism described in \cite{morton2012} 
as implemented on the publicly available \texttt{vespa}\footnote{\url{https://github.com/timothydmorton/VESPA}} package. In short, \texttt{vespa} 
considers all the false-positive scenarios that might give rise to the observed periodic dips in the light curve and, using photometric and spectroscopic 
information of the target star, calculates the false-positive probability (FPP) which is the complement of the probability of there being a planet given the observed 
signal. Because our archival and modern imaging presented on \S2.4 rule out any companion at distances larger than $9\arcsec$ radius, we consider this radius 
in our search for possible false-positive scenarios using \texttt{vespa}, which considers the area around the target star in which one 
might suspect false-positives could arise. The algorithm calculates the desired probability as

\begin{eqnarray*}
\textnormal{FPP} = \frac{1}{1+f_p P },
\end{eqnarray*}   

\noindent where $f_p$ is the occurrence rate of the observed planet (at the specific observed radius) and $P = L_\textnormal{TP}/L_\textnormal{FP}$, 
where TP indicates the transiting-planet scenario and FP the false-positive scenario, and each term is defined as $L_i = \pi_i \mathcal{L}_i$, where 
$\pi_i$ is the prior probability and $\mathcal{L}_i$ is the likelihood of the $i$-th scenario. For our target, considering all the information 
gathered and the fact that no secondary eclipse larger than $\approx 165$ ppm (i.e., 3-sigma) is detected, we obtain a value of $P = 4288.79$. 
As for the occurrence rate of planets like the one observed, we consider the rates found by \cite{petigura2013} for planets between $2-2.83R_\Earth$ 
with periods between 5 and 50 days orbiting solar-type stars, which is $7.8\%$, i.e., 
$f_p = 0.078$. This gives us a false-alarm probability of $\textnormal{FPP} = 3\times 10^{-3}$. Given that this probability is smaller than the usual $1\%$ 
threshold \cite[e.g.,][]{montet2015}, we consider our planet validated. We note that this FPP is an upper limit on the real FPP given our AO and 
lucky-imaging observations. Both observations rule out an important part of the parameter space for blending scenarios between 
$0\farcs2$ and $5\arcsec$ from the star, which are the main source of false-positives for our observations.

\subsection{Transit dilutions}

As it will be discussed in the next section, both the planet radius and mass puts BD+20594 in a very interesting part of the mass-radius diagram. Therefore, 
it is important to discuss the constraints that our spectroscopic and new, AO and lucky imaging observations pose on possible background stars that might 
dilute the transit depth and thus cause us to underestimate the transit radius.

Given that the factor by which the planetary radius is changed by a collection of stars inside the aperture used to obtain the photometry of the target 
star is given by $\sqrt{1/F_\%}$, where $F_\%$ is the fraction of the total flux in the aperture added by the star being transited, we estimate that only 
stars with magnitude differences $\lesssim 2$ are able to change the transit radius by magnitudes similar to the quoted uncertainties in Table 
\ref{table:planet-params}. We note that such magnitude differences in the Kepler bandpass are ruled out from $0\farcs2$ to the aperture radius used to obtain 
the photometry for our target star: our AO and lucky imaging observations rule out companions of such magnitudes from $0\farcs2$ to $5\arcsec$ (see Figure 
\ref{contrast-plot}). On the other hand, stars with magnitude differences of that order should be evident on our retrieved archival and new images presented 
in \S2.4, at least at distances of $5\arcsec$ from our target star, and up to and beyond the $12\arcsec$ aperture used to obtain the K2 photometry. Given that 
the remaining unexplored area on the sky is very small (only $0\farcs2$ around our target star), and that a star of such magnitude should produce an evident 
peak on the cross-correlation function on our high resolution spectra which is not seen, we consider that our derived transit radius is confidently unaffected by 
dilutions of background field stars.

\section{Discussion}
\begin{figure*}
\plotone{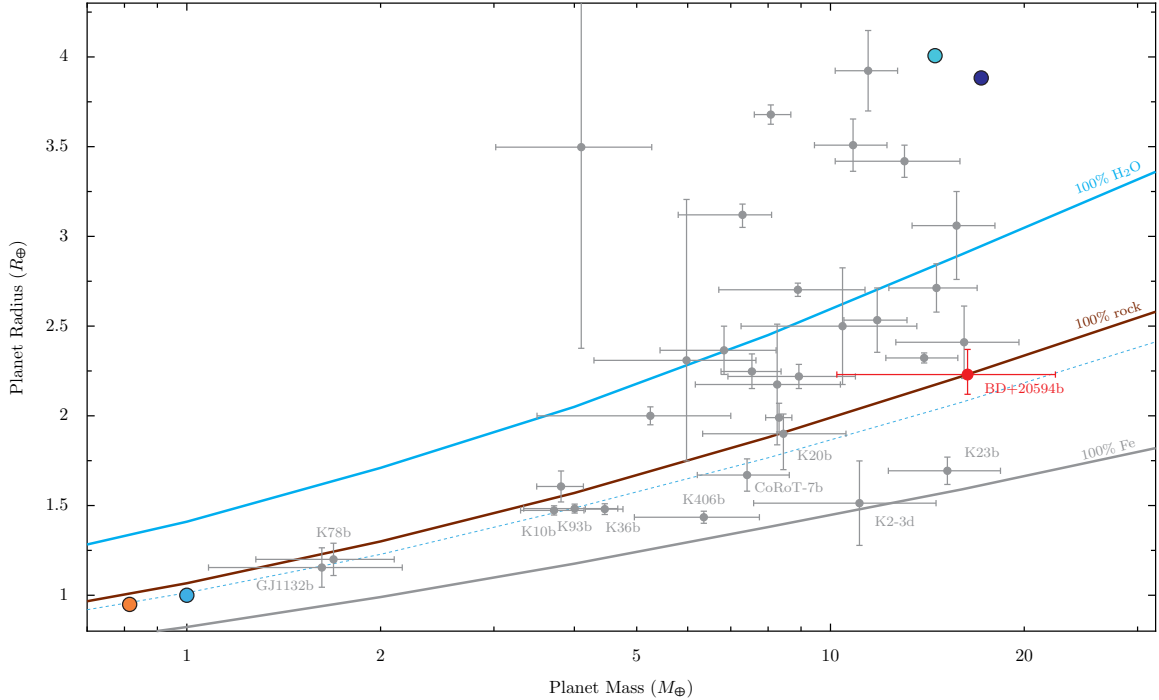}
\caption{Mass-radius relationship for planets with secure masses and radii (at the $3-\sigma$ level, grey points) having masess less than $32M_\Earth$ and radii less than 
$4R_\Earth$. BD+20594b is plotted in red, while Solar System planets are plotted as coloured circles (on the lower left Earth with blue, 
Venus with orange, and in the upper right Neptune in cyan and Uranus in dark blue). Theoretical 2-layer mass-radius models from 
\cite{zs2016} are plotted with different colors; a $100\%$ water composition is depicted in blue, a $100\%$ rock (MgSiO$_3$) composition 
in brown and a $100\%$ Fe composition in grey. The light blue dashed line indicates the best-fit composition of small rocky exoplanets 
obtained by \cite{zs2016} for reference ($74\%$ MgSiO$_3$, $26\%$ Fe); the best-fit composition of BD+20594b is that of $100\%$ MgSiO$_3$.
 \label{mr-diagram}}
\end{figure*}  

As mentioned in the previous section, the large mass ($M_p = 16.3^{+6.0}_{-6.1} M_\Earth$) for the calculated 
radius ($R_p = 2.23^{+0.14}_{-0.11}R_\Earth$) found for BD+20594b is very interesting. Figure~\ref{mr-diagram} compares 
BD+20594b with other discovered exoplanets with radii less than $4R_\Earth$ ($\sim$ Neptune) and masses 
smaller than $32M_\Earth$ (limits of theoretical models) as retrieved from exoplanets.eu\footnote{Data retrieved 
on 23/12/2015} except for the Kepler-10 planets, 
for which we use the masses obtained by \cite{weiss2016}, along with 2-layer models obtained from \cite{zs2016}. As can be seen, 
BD+20594b spans a regime in radius at which most exoplanets have low densities and are composed of large amounts of volatiles 
\citep{Rogers2015}. In particular, taking the mass-radius 
estimates for BD+20594b at face value, the best-fit composition assuming a 2-layer model for the planet is $100\%$ MgSiO$_3$, i.e., a pure 
rock composition, positioning the planet in the boundary of ``possibly rocky" and ``non-rocky" planets. More realistic three-layer alternatives, however, 
can explain the observed radius and mass of the planet if a rock/Fe core has an added volatile envelope, composed either by water or H/He 
\citep[see, e.g., the modelling for Kepler-10c in ][]{weiss2016}. If, for example, we assume an Earth-like interior composition for the planet 
(i.e., $74\%$ MgSiO$_3$ and $26\%$ Fe) and again take the mass and radius estimates at face value, three-layer models obtained from 
\cite{zs2016} give a possible $0.2R_\Earth$ water envelope for the planet (corresponding to $8\%$ in mass). This thus gives a maximum radius 
for a possible H/He envelope, which would anyways produce a small layer of much less than a percent in mass; at least significantly smaller than 
the one modelled for Kepler-10c.

\begin{figure}
\plotone{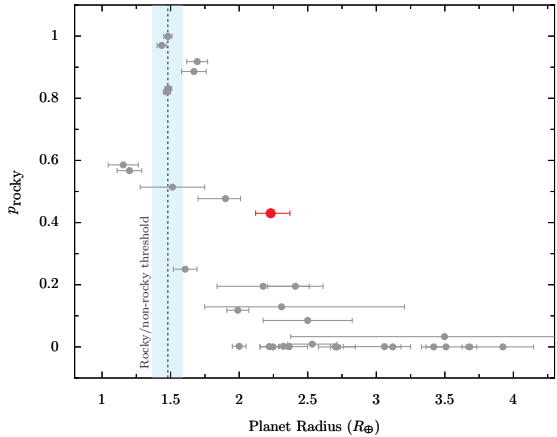}
\caption{The posterior probability that a planet is sufficiently dense to be rocky, $p_\textnormal{rocky}$, as a function of radius for all 
exoplanets with secure masses and radii (grey points), along with the estimated values for BD+20594b (red point). The black dashed 
line shows the transition between rocky (to the left) and non-rocky (to the right of the diagram) planets, along with the 95\% confidence 
band on this threshold (blue band).
 \label{procky}}
\end{figure}

Given that the errors on the mass of BD+20594b are large enough to be consistent with several compositions, a careful assessment must be made 
in order to explore its possible rocky nature. To this end, we follow 
the approach introduced by \cite{Rogers2015} and compute $p_\textnormal{rocky}$, the posterior probability that a planet is 
sufficiently dense to be rocky, which is defined as the fraction of the joint mass-radius posterior distribution that falls between a 
planet composition consistent with being rocky. A probably rocky planet, then, would have $p_\textnormal{rocky}\sim 1$, while a 
planet with a density that is too low to be rocky would result in $p_\textnormal{rocky}\sim 0$. The definition of ``rocky planet" used 
in \cite{Rogers2015}, which we adopt in this work, is given by those planets spanning compositions between $100\%$ rock and $100\%$ Fe. 
Although this definition is based on simple 2-layer models for the planetary composition, and in theory for a 
given point in the mass-radius diagram planets could have denser compositions with a gaseous envelope on top, we use 
this metric anyways in order to compare our newly discovered exoplanet in terms of the population of already discovered small 
planets. This is an important point to make, as $p_\textnormal{rocky}$ is actually an upper limit on the probability that a planet 
is indeed rocky. To compute this value and compare it to the population of exoplanets with secure masses and radii discovered so far, we use the models from 
\cite{zs2016}. To sample from the posterior distributions given the posterior estimates published in the 
literature for the different exoplanets, we use the methods described in Appendix~A of \cite{EJa2015} and assume these radii and 
masses are drawn from skew-normal distributions in order to use the asymmetric error bars published for those parameters, while 
we use the posterior samples of our MCMC fits described in \S3.2 to sample from the posterior joint distribution of mass and radius 
of BD+20594b. Our results are depicted in Figure~\ref{procky}, where we also indicate the threshold radius found by \cite{Rogers2015} 
at which there is a significant transition between rocky and non-rocky exoplanets, with smaller exoplanets having in general rocky 
compositions and larger exoplanets having less dense compositions.

As evident in Figure~\ref{procky}, BD+20594b is in an interesting position in this diagram. The closest exoplanet to BD+20594b 
in this diagram is Kepler-20b, which has a radius of $1.91^{+0.12}_{-0.21}R_\Earth$, which is only $2-\sigma$ away from the ``rocky" 
boundary. BD+20594b, on the other hand, is more than $5-\sigma$ away from it. With a value of $p_\textnormal{rocky}\sim 0.43$, 
BD+20594b is the first Neptune-sized exoplanet to date with a large (compared to the typical Neptune-sized planet) posterior probability 
of being dense enough to be rocky. 

The large mass obtained for BD+20594b implies that if the planet ever had the chance to acquire an atmosphere, it should retain it. 
However, if the planet is indeed actually primarly composed of rock, given its small radius, a significant H/He 
envelope is unlikely in the usual settings of planet formation. Calculations using core accretion theory by \cite{IH2012}, predict that if the mass 
of rock in the protoplanet is on the order of $\sim 10 M_\Earth$, even for disk dissipation time-scales on the order of $\sim 10$kyr an accretion 
of a $\sim 1M_\Earth$ H/He envelope should happen. Even in the case of a large opacity of the protoplanetary disk, a mass of rock similar to the 
one possible for BD+20594b should imply at least this level of H/He accretion. Given the bulk composition and distance of BD+20594b to its parent 
star, mass loss due to X-ray and Extreme UV radiation from its parent star its unlikely. If this indeed is the primary composition of this planet, it might be 
possible that it formed at late stages in the protoplanetary disk, under conditions similar to those on transition disks \citep{lee2016} or that some 
external effect removed the accreted envelope from the planet. Recent studies on giant impacts, which predict efficient devolatilization mechanisms for 
Super-Earths, might prove useful in explaining the lack of an extended atmosphere for BD+20594b if the planet ever accreted a significant H/He atmosphere 
in the first place \citep{SF2015}.

In terms of mass and radius, BD+20594b is similar to both Kepler-131b \citep{marcy2014} and Kepler-10c \citep{weiss2016}. 
Although both of them are probably non-rocky due to their low $p_\textnormal{rocky}$ ($\sim 0.1$ and $\sim 0.002$ respectively), which is the main 
difference with BD+20594b, they are also ``warm" Neptune-sized planets just as BD+20594b, with periods of $16d$ and $45.29d$, respectively. The 
similarity in mass, radius and period between Kepler-10c and BD+20594b, in fact, makes both of these planets excellent laboratories 
for comparison in order to put planet formation theories to test.

Finally, it is interesting to mention that the sub-solar metallicity of the host star adds more weight to the growing evidence that low-mass planets 
tend to be found orbiting stars with a lower metallicity content \citep{mayor2009,adibekyan2012} or at least they appear to show a lack of 
preference towards metal-rich stars \citep{Jenkins2013,BL2015}.

\section{Conclusions}

Using K2 photometry from Campaign 4 and a follow-up effort including radial-velocities from the HARPS spectrograph, we have presented BD+20594b, 
a planet with a radius $R_p = 2.23^{+0.14}_{-0.11}R_\Earth$ and mass of $M_p = 16.3^{+6.0}_{-6.1} M_\Earth$ orbiting a solar-type star. BD+20594b 
lies in an interesting position in the mass-radius diagram, in the boundary between "possibly rocky" and "non-rocky" planets. Given the brightness of the 
host star ($V=11.04$), BD+20594b is amenable for future follow-up studies, which will enable a detailed study of its mass and hence composition, that might 
be able to confirm whether BD+20594b is in the "possibly rocky" or "non-rocky" regime on the mass-radius diagram.

\section{Acknowledgments}
We thank the referee for insightful comments that greatly improved
this work. N.E., J.S.J. and A.J. would like to thank E. Pall\'e for his willingness to share time on northern hemisphere facilities for follow-up efforts. N.E. and R.B. 
are supported by CONICYT-PCHA/Doctorado Nacional. A.J. acknowledges support from FONDECYT project 1130857 and from BASAL CATA PFB-06. N.E., 
R.B. A.J. and J.C. acknowledge support from the Ministry for the Economy, Development, and Tourism Programa Iniciativa Cient\'ifica Milenio through 
grant IC 120009, awarded to the Millennium Institute of Astrophysics (MAS). J.S.J. acknowledges support from BASAL CATA PFB-06.
This paper includes data collected by the Kepler mission. Funding for the Kepler mission is provided by the NASA Science Mission directorate. 
It also made use of the SIMBAD database (operated at CDS, Strasbourg, France), NASA's Astrophysics Data System Bibliographic Services, and data 
products from the Two Micron All Sky Survey (2MASS) and the APASS database and the Digitized Sky Survey. Based on observations collected 
at the European Organisation for Astronomical Research in the Southern Hemisphere under ESO programmes 096.C-0499(A), 096.C-0417(A) and 096.D-0402(A).

\end{document}